RESEARCH **Open Access**

# Non-linear associations of amyloid-β with resting-state functional networks and their cognitive relevance in a large community-based cohort of cognitively normal older adults

Junjie Wu[1,2], Benjamin B. Risk[3], Taylor A. James[1], Nicholas T. Seyfried[1,4], David W. Loring[1], Felicia C. Goldstein[1], Allan I. Levey[1], James J. Lah[1*] and Deqiang Qiu[2,5*]

## Abstract

**Background** Non-linear alterations in brain network connectivity may represent early neural signatures of Alzheimer's disease (AD) pathology in cognitively normal older adults. Understanding these changes and their cognitive relevance may help clarify early network vulnerability associated with AD pathology. Most prior studies recruited participants from memory clinics, often with subjective memory concerns, limiting generalizability.

**Methods** We examined 14 large-scale functional brain networks in 968 cognitively normal older adults recruited from the community using resting-state functional MRI, cerebrospinal fluid (CSF) biomarkers (amyloid-β 1–42 [Aβ], total tau, phosphorylated tau 181), and neuropsychological assessments. Functional networks were identified using group independent component analysis.

**Results** Inverted U-shaped associations between CSF Aβ and functional connectivity were observed in the precuneus network and ventral default mode network (DMN), but not in the dorsal DMN, indicating network-specific vulnerability to early amyloid pathology. Higher connectivity in Aβ-related networks, including dorsal and ventral DMN, precuneus, and posterior salience networks, was associated with better visual memory, visuospatial, and executive performance. No significant relationships were observed between CSF tau and functional connectivity.

**Conclusions** Using a large, community-based cohort, we demonstrate that non-linear alterations in functional connectivity occur in specific networks even during the asymptomatic phase of AD. Moreover, Aβ-related network connectivity is cognitively relevant, highlighting early network vulnerability and its functional consequences in amyloid pathology.

*Correspondence:
James J. Lah
jlah@emory.edu
Deqiang Qiu
dqiu3@emory.edu
Full list of author information is available at the end of the article

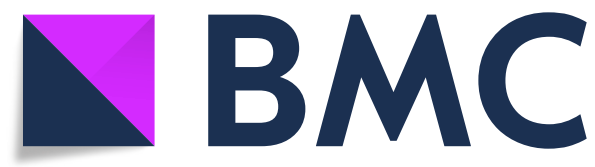

## Background

Alzheimer's disease (AD) is a progressive neurodegenerative disorder characterized by the deposition of amyloid-β (Aβ) plaques and neurofibrillary tangles of hyperphosphorylated tau (P-tau) proteins in the brain [1, 2]. Clinical symptoms of cognitive impairment can take decades to appear following the onset of AD pathology [1–3]. Identifying sensitive biomarkers of this early, asymptomatic stage is therefore critical for elucidating neural alterations associated with AD and informing the development of preventive strategies [4].

In this context, functional MRI provides a valuable approach for capturing subtle synaptic and network-level abnormalities that precede structural atrophy and cognitive decline [1, 5]. Changes in intrinsic functional connectivity have emerged as early signatures of AD pathology [6]. Cortical Aβ deposition has been associated with aberrant connectivity in vulnerable regions such as the hippocampus [7] and default mode network (DMN) [8]. Notably, these alterations often follow a biphasic trajectory, with early hyperconnectivity interpreted as compensatory responses, followed by hypoconnectivity as network failure emerges with advancing pathology [9, 10]. By contrast, tau pathology appears to exert later and detrimental effects on neural activity and connectivity [11–14].

Despite these advances, prior studies in cognitively normal older adults have important limitations. Most have targeted individual networks, such as the DMN [10, 15–20], executive control [15, 16], or salience [10] networks, typically in small samples ($N < 100$). Moreover, many cohorts were recruited from memory clinics, where participants frequently present with subjective memory concerns, raising the possibility of selection bias toward individuals already on a symptomatic trajectory. Consequently, it remains unclear whether biphasic connectivity changes extend beyond the DMN, how they differentially relate to Aβ burden versus tau pathology as indexed by cerebrospinal fluid (CSF) P-tau 181, and whether they hold measurable cognitive relevance in large community-based samples of cognitively normal older adults.

The present study addresses these gaps by systematically examining associations between CSF biomarkers of Aβ and tau and intrinsic functional connectivity across 14 large-scale brain networks in 968 cognitively normal individuals recruited from the community. We tested for nonlinear (quadratic) associations to capture potential biphasic trajectories of brain connectivity with AD biomarkers. In addition, we assessed the cognitive significance of these network alterations by correlating functional connectivity with performance on a comprehensive neuropsychological battery spanning memory, language, visuospatial, and executive domains. By integrating large sample size, community-based recruitment, biomarker quantification, and nonlinear modeling, this study provides new insights into the earliest network changes in the AD cascade and their relevance for cognition.

## Methods

### Participants

This study included 968 cognitively normal older participants (median age = 63.8 [58.1–69.2] years, 641 females [66.2%]) from Emory Healthy Brain Study [21] (Table 1). This Health Insurance Portability and Accountability Act-compliant study was approved by the institutional review board at Emory University School of Medicine. Written informed consent was obtained from all participants prior to study procedures in accordance with the Declaration of Helsinki.

### MRI acquisition

MRI data were acquired on a Siemens Magnetom Prisma 3T scanner (Siemens Healthcare, Erlangen, Germany) equipped with a 32-channel head array coil. $T_1$-weighted anatomical images were acquired using a magnetization-prepared rapid acquisition with gradient echo sequence (TR/TE = 2300/2.96 ms, TI = 900 ms, FA = 9°, voxel size = $1 \times 1 \times 1$ mm$^3$, 208 slices). A 10-min resting-state functional MRI was performed using a multiband accelerated gradient-echo echo-planar imaging sequence (TR/TE = 1890/30 ms, FA = 52°, voxel size = $1.5 \times 1.5 \times 1.5$ mm$^3$, 81 slices, multiband factor = 3, 320 volumes).

### MRI analysis

Preprocessing of functional MRI images was performed using CONN functional connectivity toolbox 22a (https://www.conn-toolbox.org). The functional images were corrected for $B_0$ field inhomogeneity, head motion, and timing of slice acquisition. The resultant images were then normalized to the Montreal Neurological Institute stereotaxic space, spatially smoothed with an 8 mm full-width-at-half-maximum Gaussian kernel, and bandpass-filtered to retain signal components with temporal frequency between 0.01 and 0.1 Hz. Head motion was confirmed to be within 2 mm of translation and 2° of rotation.

Spatially constrained independent component analysis (ICA) was performed on the functional MRI data to identify functional networks using GIFT toolbox 4.0b (https://trendscenter.org/software/gift). In brief, ICA is a data-driven blind source separation approach that decomposes functional MRI data into spatially independent

**Table 1** Demographics, cerebrospinal fluid, and neuropsychological assessments

| | **Median [Q1 - Q3] / *n* (%)** |
|---|---|
| *N* | 968 |
| Age (years) | 63.8 [58.1–69.2] |
| Sex (female) | 641 (66.2%) |
| Education (years) | 16.0 [16.0–18.0] |
| APOE ε4 alleles | |
| 0 | 604 (62.4%) |
| 1 | 254 (26.2%) |
| 2 | 37 (3.8%) |
| CSF Aβ (pg/ml) | 1197.0 [847.1–1555.2] |
| CSF T-tau (pg/ml) | 163.1 [127.9–212.5] |
| CSF P-tau (pg/ml) | 14.2 [10.9–18.5] |
| CSF P-tau/Aβ ratio | 0.011 [0.009–0.015] |
| RCFT immediate free recall | 18.0 [13.5–23.0] |
| RCFT delayed free recall | 18.0 [13.0–22.5] |
| Recognition of RCFT elements | 21.0 [19.0–22.0] |
| RCFT copy accuracy score | 33.0 [31.0–35.0] |
| JoLO | 26.0 [23.0–28.0] |
| RAVLT delayed recall | 10.0 [6.0–12.0] |
| Letter Fluency (FL) | 29.0 [24.0–34.0] |
| Animal Fluency | 21.0 [18.0–25.0] |
| TMTA | 32.0 [26.0–40.0] |
| TMTB | 65.0 [52.0–84.0] |

*APOE ε4* apolipoprotein E ε4, *CSF* cerebrospinal fluid, *Aβ* amyloid-β 1–42, *T-tau* total tau, *P-tau* tau phosphorylated at threonine 181, *RCFT* Rey Complex Figure Test, *JoLO* Judgment of Line Orientation, *RAVLT* Rey Auditory Verbal Learning Test, *TMTA* Trail Making Test Part A, *TMTB* Trail Making Test Part B components, which can be interpreted as functional networks, each with an associated time course [22]. In the spatially constrained implementation, prior spatial information is incorporated to guide component estimation toward known network patterns while allowing participant-specific variation [23]. Using network templates from a previous study [24], group ICA was applied to estimate components corresponding to 14 networks: the auditory network, language network, primary visual network, higher visual network, visuospatial network, sensorimotor network, basal ganglia network, precuneus network, anterior salience network, posterior salience network, left executive control network (LECN), right executive control network (RECN), dorsal DMN, and ventral DMN. Participant-specific network maps were then derived using regression-based back-reconstruction [25], and connectivity was averaged within the corresponding network template for each functional network.

### CSF sampling and analysis

CSF was collected in a standardized fashion applying common pre-analytical methods. Lumbar punctures were performed using a 24-g atraumatic Sprotte spinal needle (Pajunk Medical Systems, Norcross, Georgia, USA) with aspiration. After clearing any blood contamination, CSF was transferred into 15-ml polypropylene tubes (Corning, Glendale, Arizona, USA) followed by freezing in 0.5 ml aliquots on dry ice within 1 h after collection. Aliquots were stored in 0.9 ml FluidX tubes (Azenta, Chemsford, Massachusetts, USA) at -80 °C. Following a single freeze-thaw cycle, amyloid-β 1–42 (Aβ), total tau (T-tau), and tau phosphorylated at threonine 181 (P-tau) assays were performed on CSF samples on a Roche Cobas e601 analyzer using the Elecsys immunoassay platform [26]. All assays were performed in a single laboratory in the Emory Goizueta Alzheimer's Clinical Research Unit. Lower CSF Aβ 1–42 reflects greater cortical amyloid deposition, and the CSF P-tau/Aβ ratio was calculated as a combined measure of amyloid and phosphorylated tau burden. We used a CSF P-tau/Aβ ratio cutoff of 0.023 [27] to classify participants as biomarker-positive (BM+; P-tau/Aβ > 0.023) or biomarker-negative (BM-; P-tau/Aβ < 0.023).

### Neuropsychological assessments

A neuropsychological test battery was administered, including the Rey Complex Figure Test (RCFT) [28] for visual memory and visuospatial functioning, the Judgment of Line Orientation (JoLO) [29] for visuospatial ability, the Rey Auditory Verbal Learning Test (RAVLT) [28] for verbal learning and memory, the Letter Fluency (FL) [30] for language and executive functioning, the Animal Fluency [30] for language and semantic memory, the Trail Making Test Part A (TMTA) [31] for processing speed, and the Trail Making Test Part B (TMTB) [31] for executive functioning. Letter Fluency is considered an indicator of executive functioning, as it requires rule maintenance, strategic and flexible retrieval, response inhibition, and self-monitoring under time constraints [30, 32].

### Statistical analysis

Associations of functional connectivity with CSF biomarker measurements were estimated using multiple regression including quadratic terms for CSF biomarkers, with age and sex as covariates. Correlations between functional connectivity and neuropsychological performance were evaluated using multiple regression with age and sex as covariates. To evaluate whether connectivity–neuropsychological correlations varied by CSF biomarker status, we fitted additional multiple regression models including functional connectivity, biomarker group, and their interaction, with age and sex as covariates. The statistical analyses used a two-tailed level of 0.05 for defining statistical significance, and the Benjamini-Hochberg false discovery rate (FDR) procedure was applied to correct for multiple testing.

To further examine whether Aβ-related connectivity changes were cognitively relevant, mediation analyses were performed to test whether CSF Aβ was indirectly associated with neuropsychological performance through functional connectivity in the networks and outcomes exhibiting Aβ-related effects. The mediator model specified functional connectivity as the dependent variable and included linear and quadratic CSF Aβ terms, with age and sex as covariates. The outcome model specified neuropsychological performance as the dependent variable and included functional connectivity and the same linear and quadratic CSF Aβ terms, with age and sex as covariates. Because the inclusion of a quadratic Aβ term implies that both direct and indirect effects are conditional on Aβ level, derivative-based conditional direct and indirect effects were estimated across the observed CSF Aβ range using 5,000 bootstrap resamples to derive 95% percentile confidence intervals (CI). Effects were considered statistically significant when the CI did not include zero.

## Results

Table 1 summarizes participant demographic characteristics, CSF biomarker measures, and neuropsychological assessment results; corresponding information stratified by CSF biomarker status (BM- and BM+) is presented in Supplementary Table S1.

Figure 1 shows mean connectivity maps across all participants for the 14 functional networks.

Figure 2 presents correlations between CSF biomarkers and functional networks. In the precuneus network, both the linear ($\beta = 0.126$, 95% CI: 0.058–0.193, $P = 0.004$, FDR corrected) and quadratic ($\beta$ = -0.071, 95% CI: -0.122 to -0.020, $P = 0.045$, FDR corrected) terms for Aβ were significant, indicating a non-linear (inverted U-shaped) relationship between Aβ levels and functional connectivity; the combined Aβ terms ($A\beta + A\beta^2$) also improved model fit ($\Delta R^2 = 0.015$, $P = 0.008$, FDR corrected). In the ventral DMN, the quadratic term for Aβ was significant ($\beta$ = -0.089, 95% CI: -0.141 to -0.037, $P = 0.012$, FDR corrected), whereas the linear term was not ($\beta = 0.079$, 95% CI: 0.010–0.148, $P = 0.119$, FDR corrected); nevertheless, the combined Aβ terms improved model fit ($\Delta R^2 = 0.012$, $P = 0.018$, FDR corrected). These findings suggest that non-linear associations between CSF Aβ and functional connectivity are evident across multiple large-scale networks. Violin plots illustrating these non-linear trends more clearly are shown in Supplementary Fig. S1. In sensitivity analyses additionally adjusting for education and apolipoprotein E ε4 (APOE ε4), results were comparable (Supplementary Fig. S2). Specifically, in the precuneus network, both the linear ($\beta = 0.128$, 95% CI: 0.052–0.204, $P = 0.014$, FDR-corrected) and quadratic ($\beta$ = -0.082, 95% CI: -0.137 to -0.027, $P = 0.024$, FDR-corrected) Aβ terms remained significant, and the combined Aβ terms improved model fit ($\Delta R^2 = 0.014$, $P = 0.018$, FDR-corrected). In the ventral DMN, the quadratic Aβ term remained significant ($\beta$ = -0.083, 95% CI: -0.139 to -0.028, $P = 0.024$, FDR-corrected), whereas the linear term was not ($\beta = 0.061$, 95% CI: -0.016–0.139, $P = 0.522$, FDR-corrected), and the combined Aβ terms showed a trend-level improvement in model fit ($\Delta R^2 = 0.009$, $P = 0.094$, FDR-corrected). Given that APOE ε4 data were available for 895 of 968 participants (Table 1), subsequent analyses adjusted for age and sex to maximize sample size and ensure consistency across models.

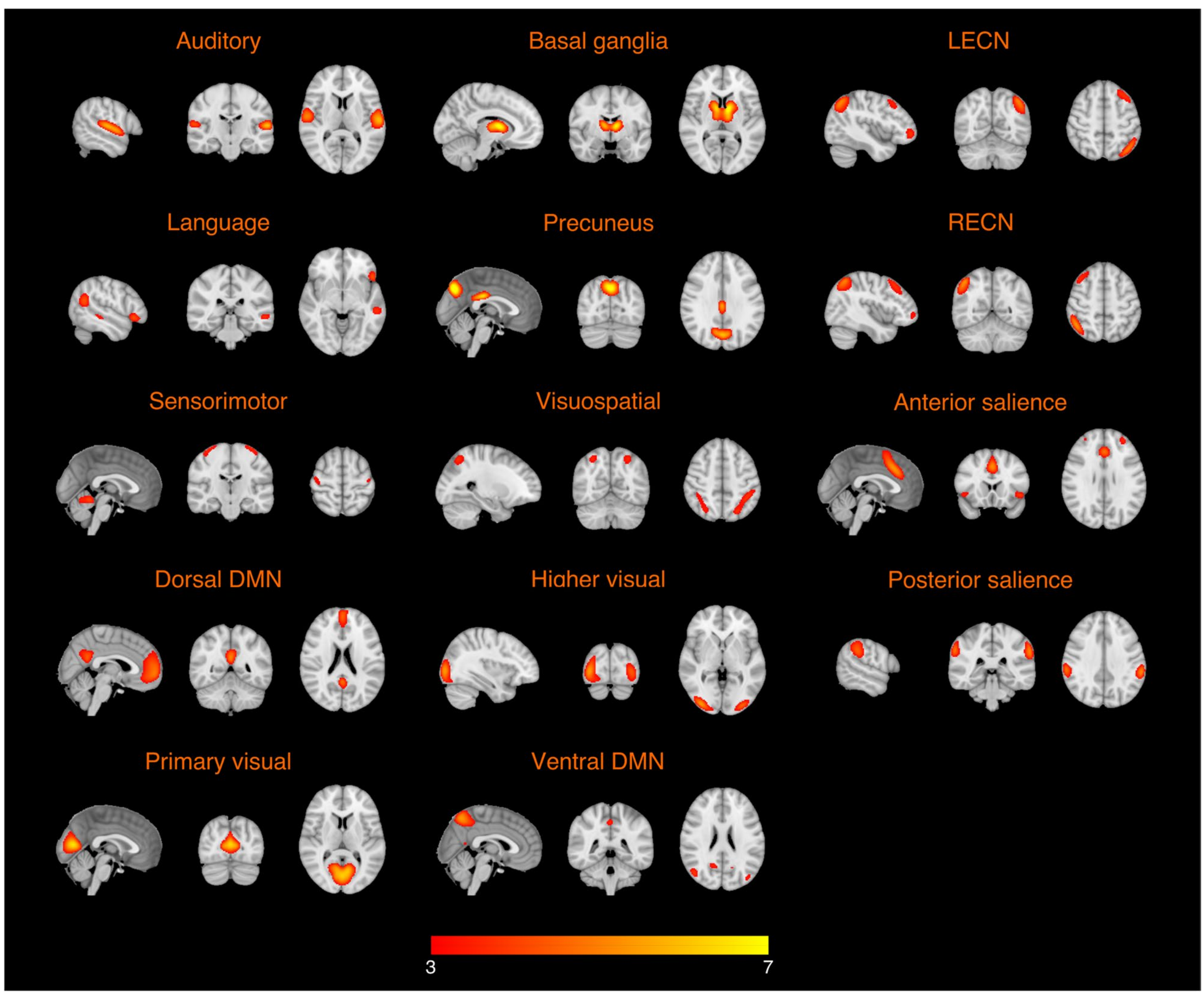

**Fig. 1** Averaged connectivity maps across all healthy older adults for the auditory network, basal ganglia network, left executive control network (LECN), language network, precuneus network, right executive control network (RECN), sensorimotor network, visuospatial network, anterior salience network, dorsal default mode network (DMN), higher visual network, posterior salience network, primary visual network, and ventral DMN

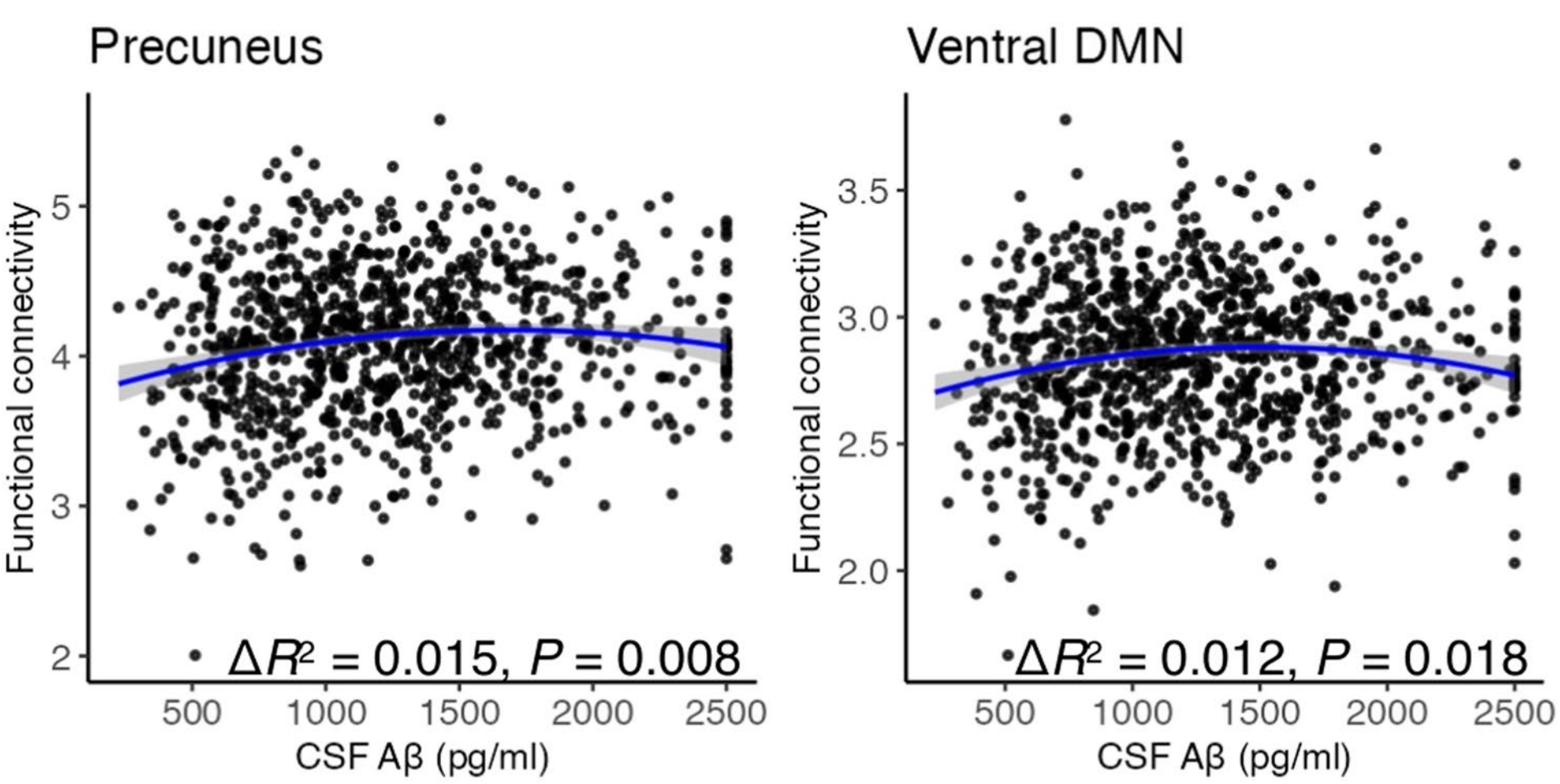

**Fig. 2** Non-linear associations between cerebrospinal fluid (CSF) amyloid-β 1–42 (Aβ) levels and functional connectivity in the precuneus network and ventral default mode network (DMN). Quadratic associations are shown by blue curves, with shaded areas representing 95% confidence intervals. Associations were evaluated using multiple regression adjusted for age and sex, with *P* values corrected using the false discovery rate

Functional connectivity was correlated with performance across multiple cognitive domains (Fig. 3). In visual memory and visuospatial functioning, higher connectivity in the precuneus ($\beta = 0.168$, 95% CI: 0.099–0.236, $\Delta R^2 = 0.027$, $P < 0.001$, FDR corrected), dorsal DMN ($\beta = 0.168$, 95% CI: 0.099–0.236, $\Delta R^2 = 0.027$, $P < 0.001$, FDR corrected), posterior salience ($\beta = 0.141$, 95% CI: 0.074–0.208, $\Delta R^2 = 0.020$, $P = 0.001$, FDR corrected), and ventral DMN ($\beta = 0.100$, 95% CI: 0.033–0.168, $\Delta R^2 = 0.010$, $P = 0.043$, FDR corrected) networks was associated with better RCFT immediate recall. RCFT delayed recall showed similar associations in the precuneus ($\beta = 0.147$, 95% CI: 0.078–0.216, $\Delta R^2 = 0.021$, $P = 0.001$, FDR corrected), dorsal DMN ($\beta = 0.147$, 95% CI: 0.078–0.216, $\Delta R^2 = 0.021$, $P = 0.001$, FDR corrected), and posterior salience ($\beta = 0.120$, 95% CI: 0.052–0.188, $\Delta R^2 = 0.014$, $P = 0.012$, FDR corrected) networks, with a trend toward significance for ventral DMN ($\beta = 0.093$, 95% CI: 0.025–0.161, $\Delta R^2 = 0.009$, $P = 0.069$, FDR corrected). Precuneus connectivity also correlated positively with RCFT copy accuracy ($\beta = 0.116$, 95% CI: 0.046–0.186, $\Delta R^2 = 0.013$, $P = 0.021$, FDR corrected). For visuospatial ability, higher connectivity in the precuneus ($\beta = 0.113$, 95% CI: 0.046–0.181, $\Delta R^2 = 0.012$, $P = 0.021$, FDR corrected) and dorsal DMN ($\beta = 0.099$, 95% CI: 0.031–0.166, $\Delta R^2 = 0.009$, $P = 0.043$, FDR corrected) was associated with better JoLO performance. For processing speed and executive functioning, faster completion times on TMTA and TMTB were linked to higher connectivity in the precuneus (TMTA: $\beta$ = -0.100, 95% CI: -0.165 to -0.036, $\Delta R^2 = 0.009$, $P = 0.033$, FDR corrected; TMTB: $\beta$ = -0.100, 95% CI: -0.164 to -0.037, $\Delta R^2 = 0.009$, $P = 0.032$, FDR corrected) and ventral DMN (TMTB: $\beta$ = -0.091, 95% CI: -0.153 to -0.028, $\Delta R^2 = 0.008$, $P = 0.043$, FDR corrected). An additional association between higher visual network connectivity and poorer RCFT recognition performance was observed (Supplementary Fig. S3). These connectivity–neuropsychological associations did not differ by CSF biomarker status, as no connectivity–biomarker group interaction remained significant after FDR correction (all unadjusted $P > 0.01$; all FDR-corrected $P > 0.05$).

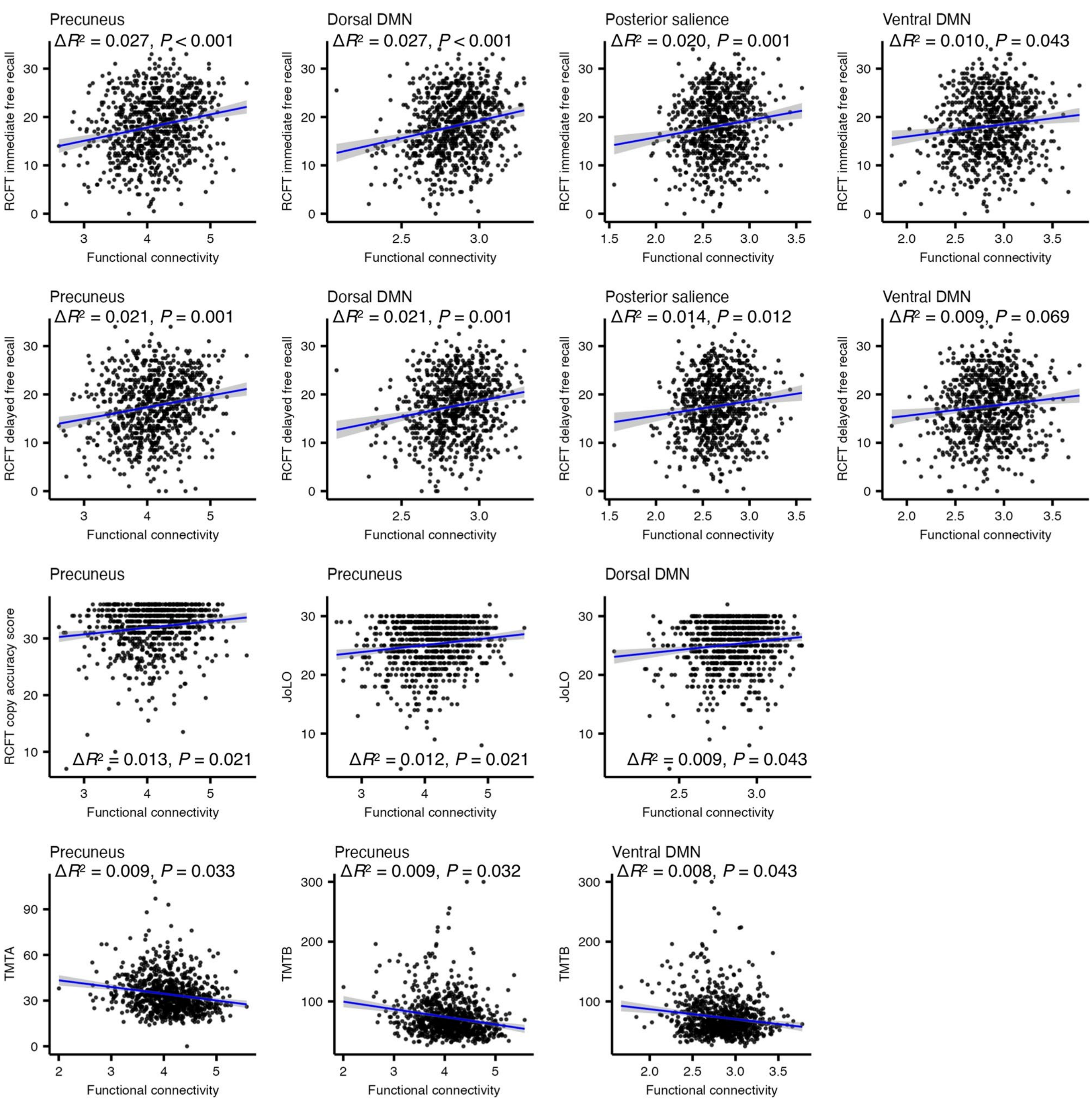

**Fig. 3** Correlations between functional connectivity and neuropsychological performance. Functional connectivity in the precuneus, dorsal default mode network (DMN), posterior salience network, and ventral DMN was positively associated with Rey Complex Figure Test (RCFT) immediate free recall. RCFT delayed free recall was positively associated with connectivity in the precuneus, dorsal DMN, and posterior salience networks, with a trend toward significance for the ventral DMN. RCFT copy accuracy was positively associated with precuneus connectivity. Visuospatial ability, measured by the Judgment of Line Orientation (JoLO), was positively associated with connectivity in the precuneus and dorsal DMN. Processing speed and executive functioning, measured by the Trail Making Test Part A (TMTA) and Part B (TMTB), were faster with higher connectivity in the precuneus and ventral DMN. Linear associations are shown by blue regression lines, with shaded areas representing 95% confidence intervals. Associations were evaluated using multiple regression adjusted for age and sex, and *P* values were corrected using the false discovery rate

Figure 4 summarizes the conditional indirect and direct effects of CSF Aβ on neuropsychological performance across the observed Aβ range. Indirect effects via functional connectivity were most evident at lower-to-intermediate CSF Aβ levels, corresponding to higher-to-intermediate cortical amyloid deposition, and attenuated toward higher CSF Aβ levels, corresponding to lower cortical amyloid deposition. In the precuneus network, indirect effects were evident at lower CSF Aβ levels for RCFT immediate and delayed free recall, RCFT copy accuracy, and JoLO, and diminished at higher CSF Aβ levels; the percentage mediated was generally higher at lower CSF

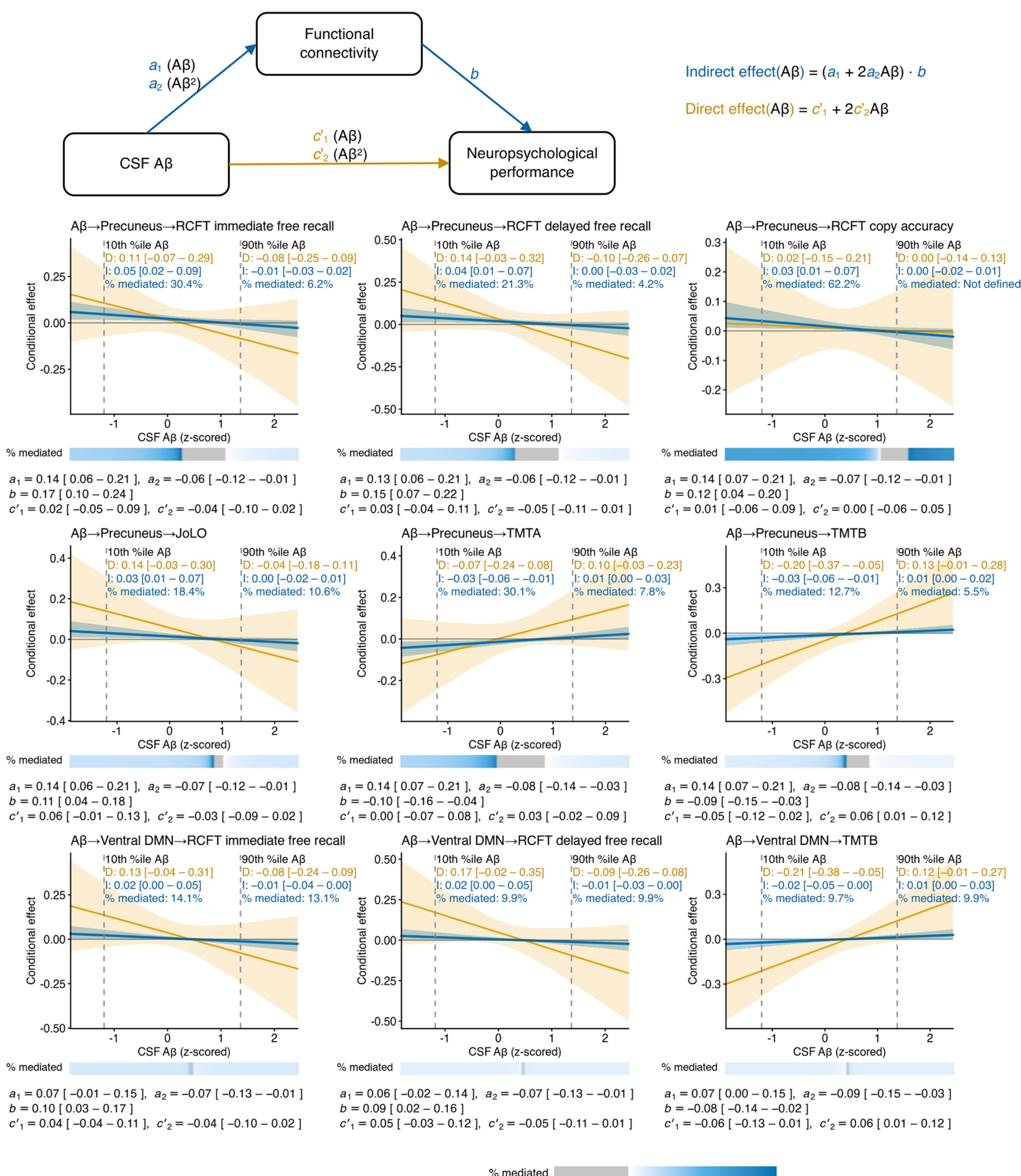

**Fig. 4** Conditional direct and indirect effects of cerebrospinal fluid (CSF) Aβ on neuropsychological performance through functional connectivity. The mediation path diagram at the top summarizes the fitted models with linear and quadratic Aβ terms (Aβ and Aβ²) for the Aβ–connectivity path ($a_1$, $a_2$), the connectivity–performance path ($b$), and the direct Aβ–performance path ($c'_1$, $c'_2$). The 3×3 plots below show conditional direct (D; orange curves) and indirect (I; blue curves) effects across the observed CSF Aβ range; shaded ribbons indicate bootstrap 95% confidence intervals. Effects were considered statistically significant when the confidence interval excluded zero. Dashed vertical lines mark the 10th and 90th percentiles of CSF Aβ. A heat strip beneath each plot summarizes percentage mediated across Aβ; gray indicates percentage mediated was not defined because direct and indirect effects had opposite signs or the total effect was approximately zero

Aβ and small at higher CSF Aβ. For TMTA and TMTB, in which lower completion times indicate better performance, indirect effects via precuneus connectivity were observed at lower and higher CSF Aβ levels with opposing directions, consistent with a quadratic association between Aβ and connectivity; the percentage mediated was small at higher CSF Aβ. In the ventral DMN, indirect effects for RCFT immediate and delayed free recall were primarily observed at lower CSF Aβ levels, whereas the indirect effect for TMTB was evident at both lower and higher CSF Aβ levels; the percentage mediated was small across the Aβ range.

## Discussion

In this large, community-based cohort of cognitively normal older adults, we provide novel insights into the earliest network changes in the AD cascade and their cognitive relevance. Specifically, inverted U-shaped associations between CSF Aβ and functional connectivity were observed in the precuneus network and ventral DMN, but not in the dorsal DMN. Higher connectivity in Aβ-related networks, including the dorsal and ventral DMN, precuneus, and posterior salience networks, was associated with better visual memory, visuospatial, and executive performance. Mediation analyses further indicated that Aβ-related effects on cognition via functional connectivity were most evident at lower-to-intermediate CSF Aβ levels, reflecting higher-to-intermediate cortical amyloid deposition, and attenuated at higher CSF Aβ levels.

Resting-state functional MRI measures are inherently variable across individuals and sessions [33], and we likewise observed wide dispersion of functional connectivity values across CSF Aβ levels. Accordingly, our findings are best interpreted as group-level associations rather than as evidence for near-term individual diagnostic utility. The incremental variance explained by Aβ terms was modest ($\Delta R^2 \approx 0.012$–$0.015$), as expected in heterogeneous asymptomatic community samples. Nevertheless, intrinsic functional connectivity provides a systems-level index of early synaptic and network dysfunction that may precede structural atrophy and cognitive decline [1, 5]. The observed non-linear Aβ–connectivity relationships and their cognitive correlates therefore provide evidence linking early amyloid burden to network-specific alterations with functional consequences.

AD is increasingly recognized to involve large-scale brain network dysfunction [1, 5, 34]. Converging evidence indicates that Aβ preferentially accumulates in the DMN and salience network [10, 35] and is associated with disrupted connectivity [10, 16]. In line with prior reports linking Aβ to non-linear alterations in DMN connectivity [9, 10], our results refine this pattern by demonstrating that such associations are evident in the precuneus network and ventral DMN, but not in the dorsal DMN. This network-specific vulnerability supports a staged model in which early Aβ deposition in high-demand DMN hubs, particularly the precuneus and the medial temporal subsystem, initially manifests as hyperconnectivity, potentially reflecting compensatory upregulation to maintain cognitive function in the presence of subtle synaptic inefficiency [10]. As Aβ burden increases, this compensation may become unsustainable, progressing toward synaptic dysfunction, neuronal loss, and eventual network breakdown, yielding the observed inverted U-shaped trajectory [36].

The precuneus and ventral DMN, encompassing hippocampal and medial temporal components, are plausible early targets [8, 37] given their central roles in episodic memory [38] and high baseline metabolic activity, which may heighten susceptibility to Aβ-related disruption [39]. In contrast, the dorsal DMN, centered on dorsal medial prefrontal regions involved in self-referential processing [40], may be less sensitive at this stage due to relatively lower metabolic demand or later involvement in the disease cascade [8, 37]. However, methodological factors related to network definition may also influence sensitivity across DMN subsystems. In spatially constrained ICA, template-guided boundaries and regional contributions within ventral and dorsal DMN templates can shape component estimation, and ventral DMN signals may be more affected by inter-individual anatomical variability and susceptibility-related signal loss in inferior and medial temporal regions. Nonetheless, the use of the same validated templates and identical preprocessing and ICA procedures across participants mitigates systematic bias, making it unlikely that network definition alone explains the observed ventral–dorsal dissociation. Together, these findings underscore the importance of examining DMN subsystems separately rather than treating the DMN as a unitary network when characterizing early Aβ-related network alterations.

We did not detect significant associations between global CSF tau measures and functional connectivity in any network. Prior work, particularly tau PET studies quantifying regional fibrillar tau burden, has demonstrated that tau accumulation follows large-scale functional networks and is closely linked to network dysfunction, with effects that strengthen as tau burden increases and disease progression advances [11, 41–47]. In our cognitively normal, community-based sample with relatively modest tau pathology, CSF tau measures may be less sensitive to network-specific effects because they do not capture the regional distribution of aggregated tau. Moreover, CSF P-tau 181 predominantly reflects Aβ-driven tau phosphorylation [48], which may obscure independent effects of tau on functional connectivity in the asymptomatic phase. Future studies combining resting-state functional MRI with tau PET will be critical to determine when and where tau pathology exerts detectable effects on large-scale functional networks in cognitively normal older adults.

AD-related cognitive impairment is reported in diverse domains, including memory, attention, executive function, language and visuospatial ability [35]. In the present study, higher connectivity within Aβ-related brain networks [10, 35], i.e., dorsal and ventral DMN, precuneus, and posterior salience networks, was associated with better neuropsychological performance, particularly in visual memory, visuospatial, and executive domains. These findings suggest that even subtle network disruptions during the asymptomatic phase of AD may have measurable cognitive relevance.

Several limitations should be noted. First, our sample included a disproportionately higher number of female participants, reflecting women's greater willingness to participate in research. This imbalance may limit representativeness, and future studies should strive for a more balanced sex distribution to improve generalizability. Second, while this cross-sectional study benefits from a large sample size, longitudinal evaluation of changes in brain network connectivity is necessary to better understand the temporal dynamics of network alterations. Follow-up data are currently being collected as part of the ongoing Emory Healthy Brain Study.

## Conclusions

In conclusion, this study provides novel evidence that early Aβ pathology is associated with non-linear alterations in functional connectivity within the precuneus network and ventral DMN, but not the dorsal DMN. Aβ-related brain networks, including the dorsal and ventral DMN, precuneus, and posterior salience networks, support visual memory, visuospatial, and executive performance. Together, these findings support early network vulnerability and its functional consequences in amyloid pathology during the asymptomatic phase of AD.

### Abbreviations

| | |
|---|---|
| AD | Alzheimer's disease |
| Aβ | Amyloid-β |
| P-tau | Hyperphosphorylated tau |
| DMN | Default mode network |
| CSF | Cerebrospinal fluid |
| ICA | Independent component analysis |
| LECN | Left executive control network |
| RECN | Right executive control network |
| T-tau | Total tau |
| BM+ | Biomarker-positive |
| BM- | Biomarker-negative |
| RCFT | Rey Complex Figure Test |
| JoLO | Judgment of Line Orientation |
| RAVLT | Rey Auditory Verbal Learning Test |
| TMTA | Trail Making Test Part A |
| TMTB | Trail Making Test Part B |
| FDR | False discovery rate |
| CI | Confidence interval |
| APOE ε4 | Apolipoprotein E ε4 |

## Supplementary Information

The online version contains supplementary material available at https://doi.org/10.1186/s13195-026-01986-w.

Supplementary Material 1.

### Authors' contributions

J.W., N.T.S., D.W.L., F.C.G., A.I.L., J.J.L., and D.Q. conceptualized the study and interpreted the data. J.W. and T.A.J. performed the acquisition of the data. J.W., B.B.R., and T.A.J. performed the analysis and quality control of the image data. J.W., J.J.L., and D.Q. wrote the manuscript and all authors contributed to the reviewing and editing the manuscript.

### Funding

The study was supported by the National Institutes of Health grants/awards (R01AG089806, R01AG070937, R01AG072603, and P30AG066511).

### Data availability

The data that support the findings of this study are available upon reasonable request from qualified investigators, adhering to ethical guidelines and signing a data use agreement with the authors' institution.

## Declarations

### Ethics approval and consent to participate

This Health Insurance Portability and Accountability Act-compliant study was approved by the institutional review board at Emory University School of Medicine. Written informed consent was obtained from all participants prior to study procedures in accordance with the Declaration of Helsinki.

### Consent for publication

Not applicable.

### Competing interests

The authors declare no competing interests.

### Author details

[1]Department of Neurology, Emory University, 6 Executive Park Dr NE, Atlanta, GA 30329, USA
[2]Department of Radiology and Imaging Sciences, Emory University, 1364 Clifton Rd NE, Atlanta, GA 30322, USA
[3]Department of Biostatistics and Bioinformatics, Emory University, Atlanta, GA, USA
[4]Department of Biochemistry, Emory University, Atlanta, GA, USA
[5]Joint Department of Biomedical Engineering, Emory University, Georgia Institute of Technology, Atlanta, GA, USA

## Publisher's Note